# Design of experiments for Quick U-building (QUB) method for building energy performance measurement

Christian Ghiaus[a*], Florent Alzetto[b]

[a]Univ Lyon, CNRS, INSA-Lyon, Université Claude Bernard Lyon 1, CETHIL, UMR5008, F-69621, Villeurbanne, France
[b]Saint-Gobain Recherche, 39 quai Lucien Lefranc, Aubervilliers Cedex 93303 France

## Abstract

Quick U-building (QUB) is a method for short time measurement of energy performance of buildings, typically one night. It uses the indoor air temperature response to power delivered to the indoor air by electric heaters. This paper introduces a method for estimating the expected measurement error as a function of the amplitude and the time duration of the input signal based on the decomposition of the time response of a state-space model into a sum of exponentials by using the eigenvalues of the state matrix. It is shown that the buildings have a group of dominant time constants, which gives an exponential response, and many very short and very large time constants, which have a small influence on the response. The analysis of the eigenvalues demonstrates that the QUB experiment may be done in a rather short time as compared with the largest time constant of the building.

[*] Corresponding author.
E-mail address: christian.ghiaus@insa-lyon.fr

## Nomenclature

### Latin letters

| | |
|---|---|
| $c$ | - specific heat capacity [J/(kg K)] |
| diag(**matrix**) | - vector containing the diagonal of a matrix |
| **diag**(vector) | - diagonal matrix obtained from a vector |
| $m$ | - mass [kg] |
| $t_{QUB}$ | - duration of heating and duration of cooling in QUB experiment [s] |
| | |
| $A$ | - area [m$^2$] |
| $C$ | - heat capacity [J/K] |
| $G$ | - thermal conductance [W/K] |
| $G(s)$ | - dynamic gain of a transfer function |
| $H'$, $F_S$ | - areal heat transfer coefficient [W/(m$^2$ K)] |
| $H$ | - building heat transfer coefficient or building overall thermal conductance [W/ K] |
| $K$ | - static gain of a transfer function |
| $P_0$ | - total power delivered to the indoor air before the beginning of QUB experiment [W] |
| $P_h$ | - power delivered to the indoor air during the heating phase [W] |
| $P_c$ | - power delivered to the indoor air during the cooling phase [W] |
| $\dot{Q}$ | - heat flow rate [W] |
| $R$ | - thermal resistance [ K/W] |
| $T_o$ | - outdoor air temperature [°C] |
| $U$ | - thermal transmittance of thermal conductance [W/(m$^2$ K)] |
| $UA$ | - overall thermal conductance [W/K] |

### Greek letters

| | |
|---|---|
| $\alpha_h$ | - slope of the temperature variation during heating [°C/s] |
| $\alpha_c$ | - slope of the temperature variation during cooling [°C/s] |
| $\varepsilon_m$ | - probable error from measurements [W/ K] |
| $\varepsilon_{QUB}$ | - intrinsic error of QUB method [W/ K] |
| $\varepsilon_{QUB\%}$ | - relative intrinsic error of QUB method [–] |
| $\theta_i$ | - indoor air temperature of zone $i$ [°C] |
| $\tau$ | - time constant [s] |
| | |
| $\Delta T^c$ | - difference between the indoor and outdoor temperatures during cooling [°C] |
| $\Delta T_0^c$ | - difference between the indoor and outdoor temperatures during cooling at time $t_0$ [°C] |
| $\Delta T^h$ | - difference between the indoor and outdoor temperatures during heating [°C] |
| $\Delta T_0^h$ | - difference between the indoor and outdoor temperatures during heating at time $t_0$ [°C] |

### Vectors and matrices

| | |
|---|---|
| **f** | - vector of sources of heat flow rate |
| **b** | - vector of sources of temperature |
| **u** | - vector of inputs |
| **x** | - vector of states |
| | |
| **A** | - incidence matrix in thermal circuit; state matrix in state-space model |
| **B** | - input matrix |
| **C** | - output matrix in state-space model |
| **D** | - feedthrough matrix |
| **G** | - matrix of thermal conductances in thermal circuit |
| **K** | - matrix of coefficients |
| **V** | - matrix of eigenvectors |
| | |
| **θ** | - vector of temperatures |
| **Φ** | - state transition matrix |
| **Λ** | - matrix of eigenvalues |

# 1 Introduction

The areal heat transfer coefficient is an intrinsic indicator of the energy performance of buildings in steady-state. Therefore, the measurement of the heat transfer coefficient of residential buildings, especially in a short period of time (less than a day), is of particular interest.

The thermal transmittance of a flat wall (or the U-value) is defined as the "heat flow rate in the steady state, $\dot{Q}$, divided by area, $A$, and by the temperature difference between the surroundings of both sides of a flat uniform system, $\Delta T$ (ISO 7345, 2018):

$$U = \frac{\dot{Q}}{A\,\Delta T} \qquad (1)$$

Heat transfer coefficients are defined for the heat transmission to external environment, to adjacent spaces, and to the ground (ISO 13790, 2008; ISO 52016-1, 2017; ISO 13789, 2017). In steady-state, the U-value of the building elements can be obtained by dividing the heat flow rate over the temperature difference on the two sides of a wall (ISO 9869-1, 2014; ASTM C 1046-95, 2001; ASTM C 1155-95, 2001; Janssens, 2016). This method is valid if:
- the thermal properties of the materials and the heat transfer coefficients do not change during the measurement;
- the variation of the heat stored in the wall is negligible as compared with the heat transferred through the wall (Thébault & Bouchié, 2018).

Similarly, the areal heat transfer coefficient, $H'$, is defined as "the heat flow rate between the building (i.e. the internal environment) and the external environment, $\dot{Q}$, divided by the area, $A$, and the difference of temperature between the internal and external environment, $\Delta T$" (ISO 7345, 2018):

$$F_S \equiv H' = \frac{\dot{Q}}{A\,\Delta T} \qquad (2)$$

Therefore, the heat transfer coefficient of the building is:

$$H = \frac{\dot{Q}}{\Delta T} \qquad (3)$$

Since steady-state conditions are never achieved for whole buildings, the methods used for measuring the overall thermal transmittance are based on one or a combination of the following procedures (Deconinck & Roels, 2016):
- imposing steady-state, e.g. hot-box method for a whole wall (ISO 8990, 1994);
- considering mean values over a long period of time, longer than three days and multiple of 24h (ISO 9869-1, 2014);
- correcting the steady state method for storage effects (ISO 9869-2, canceled);

- identifying model parameters by using long-term experiments. (Fels, 1986) (Fels, 1986) Steady-state or dynamic models are used for parameter identification in long term experiments, typically a season or a whole year. In the first approach, the areal heat transfer coefficient is determined from the energy signature which represents the relation between the consumption and the temperature difference between indoors and outdoors. Basically, the areal heat transfer coefficients are determined by regression (Fels, 1986; Hitchin & Knight, 2012; Vesterberg, Andersson, & Olofsson, 2016; Danov, Carbonell, Cipriano, & Marti-Herrero, 2013; Ghiaus, 2006) or robust regression (Ghiaus, 2006). In the second approach, the areal heat transfer coefficient is estimated from the coefficients of dynamic models (Jimenez & Madsen, 2008; Roels, Bacher, Bauwens, Castano, & Jimenez, 2017; Naveros, Bacher, Ruiz, Jimenez, & Madsen, 2014; Raillon & Ghiaus, 2018; Janssens, 2016; Naveros, Ghiaus, Ruiz, & Castano, 2015).

Since measuring the performance in quasi steady-state requires typically a few weeks, shorter-time methods for unoccupied houses were investigated. In these methods, the measurements are done when the electrical appliances and other internal heat gains are zero, with the exception of the electrical energy used for the experiment, which is counted. A widely used method for measuring the overall heat transfer coefficient of a house is co-heating. The heat transfer coefficient of the whole building is determined, including the air renewal rate, by steady-state measurement and regression analysis (Bauwens & Roels, 2014). The monitoring is done typically for 5 – 10 days with measurements at a time steps of 40 – 60 min. Variants include measurement of losses through air infiltration, solar aperture, and ground floor.

PSTAR is a unified method for building simulation and short-time tests (Subbarao, Burch, Hancook, Lekov, & Balcomb, 1988). It uses energy balance equations, one-time measurements, and hourly measured data in order to obtain a description of the building at macrolevel. The main idea is to obtain a quasi-steady state by maintaining the temperature constant for several hours through the night in order to reduce at insignificancy the effect of thermal storage. The tests are done for about 36 hours during two nights, one night for co-heating and another one for cooling down. The thermal conductance is measured during the last two hours. The method uses the measurement of the air-infiltration rate. The main issue with this approach is that it assumes that steady-state is obtained in a rather short time (one night).

Another class of methods is based on finding the parameters of the models from the response to excitation pulses. ISABELE method uses least squares identification of the parameters of a first order dynamic model obtained from the response to a step cooling (Bouchié R. , et al., 2015; Bouchié R. , Alzetto, Brun, Boisson, & Thebault, 2014; Thébault & Bouchié, 2018). The experimental protocol has three steps:
- no heating power injected in order to assess the initial thermal energy stored in the building; if the building has quasi-constant indoor temperature for several weeks, this step is not required;
- heat the building for at least two days to homogenous internal temperature having values at least 10 °C larger than the temperature of the first step;
- turn off the heater for at least two days and measure the free-running temperature variation in time.

The response factors method measures the global resistance of a building element by identifying the time series parameters of the response to a rectangular or triangular pulse (Rasooli, Itard, & Ferreira, 2016; Brisken & Mitalas, 1956). The triangular shape of the excitation in heating was obtained by heating up to 70 – 90 °C with a radiative heater for 15 min; in cooling, the radiative heater was removed, and a fan and an ice bag were used. A main requirement is that the temperature on the outdoor side of the wall remains constant; therefore, the outdoor wall is insulated from the exterior conditions by using a box in which the temperature is controlled (Rasooli, Itard, & Ferreira, 2016).

QUB method consists on a positive step input followed by a zero step, i.e. the excitation is rectangular. The outdoor temperature is considered constant during the experiment. The parameters of a first order dynamic model are deduced from these measurements. This method measures the heat transfer coefficient, $H$, of a building in a rather short time, usually one night. QUB method was tested on many experimental and real houses. Some of these results have been already published:
- on a house constructed inside an environmentally controlled chamber at University of Salford (Meulemans, Alzetto, Farmer, & Gorse, 2017; Alzetto, Pandraud, Fitton, & Heusler, 2018; Alzetto, Farmer, Fitton, Huges, & Swan, 2018);
- on one of the twin houses of Fraunhofer Institute for Building Physics (Alzetto, Pandraud, Fitton, & Heusler, 2018);
- on a house of 280 m$^2$ in southern France (Mangematin, Pandraud, & Roux, 2012);
- on an apartment in a high-raise building in Stockholm (Meulemans J. , 2018).

The response factors, ISABELE and QUB methods have similarities in the experimental procedure: short time experiment, rectangular or triangular excitation, constant outdoor temperature. However, there are important differences. In the measurement procedure, response factors are designed for a building element (wall) and considers the heat flow rate and surface temperatures while ISABELE and QUB methods are designed for the whole building and consider the equivalent indoor air temperature. In the method used for interpreting the results, response factors method uses experimental values to find out the parameters of the time series, ISABELE identifies the parameters of a model with five resistances and one capacitance, while QUB method uses slopes to find the parameters of an exponential response.

A common issue not treated in these methods is the design of experiments: what is the amplitude and the duration of the excitation to be used in order to obtain minimum experimental errors. This paper responds to this question for QUB method. A second issue is the justification of relatively short time experiment as compared with the largest time constants. This paper treats this aspect by analysing of the eigenvalues of a detailed simulation model.

## 2 Principle of QUB method
The principle of QUB method is to apply two power pulses to the building and to determine the global conductivity and effective capacity from these measurements by finding the parameters of a first order dynamic model (Mangematin, Pandraud, & Roux, 2012) (Figure 1). The model on which the measurements are fitted is:

$$c\frac{d\theta_i}{dt} = P - H_{QUB}\Delta T \tag{4}$$

where
$\theta_i$     - building temperature;
$\Delta T$     - inside – outside temperature difference, $\Delta T \equiv \theta_i - T_o$;
$P$     - power delivered by electric heaters to the indoor air;
$C$     - effective heat capacity;
$H_{QUB}$     - heat transfer coefficient.

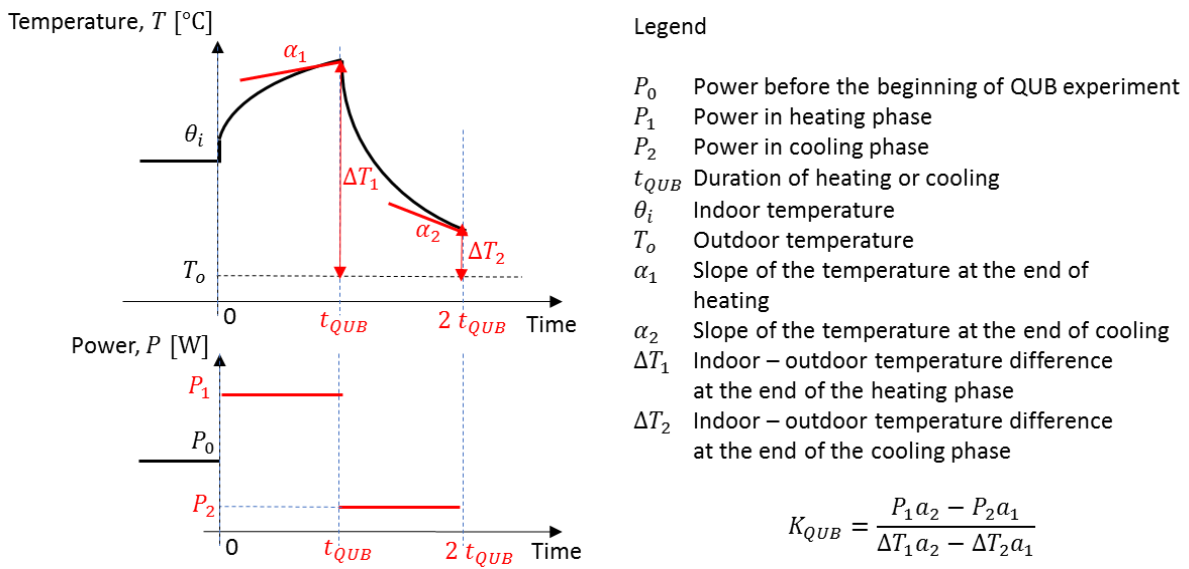

Figure 1 Principle of QUB method (Alzetto, Gossard, & Pandraud, 2014)

The two experiments conducted in QUB method are:
1) Heating: all the rooms are heated at roughly the same temperature with controlled electrical heaters; the total injected power is $P = P_h$.
2) Free-cooling: the building is left to cool. The power during the cooling period, $P = P_c$, is from interruptible sources, such as internet connection and measurement devices (Alzetto, Pandraud, Fitton, & Heusler, 2018).

In QUB method, the two parameters, $H_{QUB}$ and $C$, of the model (4) are obtained from the two slopes, $\alpha_c$ and $\alpha_h$ given by eq. (62) and eq. (55), respectively:

$$H_{QUB} = \frac{P_h \alpha_c - P_c \alpha_h}{\Delta T_0^h \alpha_c - \Delta T_0^c \alpha_h} \tag{5}$$

and

$$C = \frac{P_h \Delta T_0^c - P_c \Delta T_0^h}{\alpha_h \Delta T_0^c - \alpha_c \Delta T_0^h} \tag{6}$$

When the slopes are taken after time $t_0$, their expressions are given by (see Annex 1)

$$\alpha_h = e^{-\frac{t_0}{\tau}}(P/C - \Delta T_0^h H_{QUB}/C) \tag{7}$$

and

$$\alpha_c = e^{-t_0 \frac{H_{QUB}}{C}} \left( \frac{P_c}{C} - \frac{\Delta T_0^c H_{QUB}}{C} \right) \tag{8}$$

Therefore

$$H_{QUB}^* = \frac{P_h \alpha_c - P_c \alpha_h}{\Delta T_0^h \alpha_c - \Delta T_0^c \alpha_h} = H_{QUB} \tag{9}$$

and

$$C^* = \exp(-t_0/\tau) \frac{P_h \Delta T_0^c - P_c \Delta T_0^h}{\alpha_h \Delta T_0^c - \alpha_c \Delta T_0^h} = \exp(-t_0/\tau) C \tag{10}$$

Note that $H_{QUB}$ has the same expression regardless of the time at which the slope is considered, in origin or at time $t_0 > 0$, while the value of $C$ depends on the time at which the slope is calculated. In order to find the heat capacity $C^*$ when slopes at time $t_0 > 0$ are considered, the transcendental equation (10) needs to be solved. Equation (9) is important because it shows that the overall heat transfer coefficient, $H_{QUB}$, can be estimated based on the slopes measured at moments different from the initial time. As shown hereafter, the buildings have a dominant time constant, which gives an exponential response, and many short and large time constants, which have a small influence on the response. If the measurement is done after a time which corresponds to the steady-state for the short time constants, the time response is an exponential. Therefore, QUB method allows us to measure the global conductance by measuring the slopes of this exponential response at moments different of the initial time, long enough to settle the response due to short time constants and short enough to not be influenced by the very large time constants.

## 3  Optimal design of experiments for QUB method

Knowing the outdoor temperature $T_o$ and the power in cooling phase $P_c$, the aim of the optimal design of experiments for QUB method is to find the heating power $P_h$ and the heating time $t_{QUB}$ which minimize the total error of the measured heat transfer coefficient, $H_{QUB}$. The total error is composed by the intrinsic error of the method and the measurement error. The proposed procedure is:
- From a thermal network of the building obtain its state-space model.
- From the state-space model, calculate the overall conductance $H \equiv R^{-1}$ (an exact value) that will be considered as reference.

- Introduce the boundary conditions (temperatures and heat flows) and initial conditions and calculate $H_{QUB}$ by simulating the QUB experiment on the detailed model.
- Draw a contour map of the error $\varepsilon_{QUB} = \frac{H_{QUB} - H}{H}$ as a function of the heating power $P_h$, the time duration $t_{QUB}$, and the measurement errors.
- Chose an experiment which minimizes the error and respects technological constraints like maximum available power and admissible temperature.
- Compare the experimental response of the building with the response of the model.

## 4 Total error

The total error depends on the intrinsic error of the method and on the error coming from measurements. The intrinsic error coming from QUB method is

$$\varepsilon_{QUB} = H_{QUB} - H \qquad (11)$$

where $H_{QUB}$ is the overall heat transfer coefficient calculated by QUB experiment (Figure 1, equation (5)), and $H$ is the overall heat transfer coefficient between outdoor and indoor air, calculated from a model of the building.

The measurements needed for $H_{QUB}$ are affected by errors. In order to estimate the error from measurements, let us consider a function of several variables $f(x_1, \ldots, x_n)$. We want to evaluate the error of the function when the approximate values $a_1, \ldots, a_n$ are known and the exact values $x_1, \ldots, x_n$ are unknown.

If:
- the approximate measured values $a_1, \ldots, a_n$ are statistically independent;
- the approximate values have the errors normally distributed;
- the errors $\sigma_1 = a_1 - x_1, \ldots, \sigma_n = a_n - x_n$ are the standard deviations;
- the errors are small;

then the error (i.e. the standard deviation) of the function is (Taylor, 1997; JCGM, 2008):

$$\sigma_f = f(a_1, \ldots, a_n) - f(x_1, \ldots, x_n) = \sqrt{\left(\sigma_1 \frac{\partial f}{\partial x_1}\right)^2 + \cdots + \left(\sigma_n \frac{\partial f}{\partial x_n}\right)^2} \qquad (12)$$

The estimation of the error propagation in equation (5), needs estimation of measurement errors. The errors $\varepsilon_{Ph} = P_h - \bar{P}_h$; $\varepsilon_{Pc} = P_c - \bar{P}_c$, $\varepsilon_{\Delta Th} = \Delta T_h - \Delta \bar{T}_h$, $\varepsilon_{\Delta Tc} = \Delta T_c - \Delta \bar{T}_c$ are obtained from the characteristics of the power devices and temperature sensors. The errors on slope, $\varepsilon_{ah} = \alpha_h - \bar{\alpha}_h$; $\varepsilon_{ac} = \alpha_c - \bar{\alpha}_c$, are estimated from the correlation coefficient obtained when the linear regression is done.

Let us consider that $H$ is the exact value and $H_{QUB}$ is the measured value, which depends on errors on $\alpha_h$, $\alpha_c$, $P_h$, $P_c$, $\Delta T_h$ and $\Delta T_c$ (Figure 1).

We will note:

- $\alpha_h$, $\alpha_c$, $P_h$, $P_c$, $\Delta T_h$ and $\Delta T_c$ the measured values that are affected by absolute errors which are relatively small as compared to the true values.
- $\bar{\alpha}_h$, $\bar{\alpha}_c$, $\bar{P}_h$, $\bar{P}_c$, $\Delta \bar{T}_h$ and $\Delta \bar{T}_c$ the true (exact) values.
- $\varepsilon_{ah} = \alpha_h - \bar{\alpha}_h$; $\varepsilon_{ac} = \alpha_c - \bar{\alpha}_c$; $\varepsilon_{Ph} = P_h - \bar{P}_h$; $\varepsilon_{Pc} = P_c - \bar{P}_c$, $\varepsilon_{\Delta Th} = \Delta T_h - \Delta \bar{T}_h$, $\varepsilon_{\Delta Tc} = \Delta T_c - \Delta \bar{T}_c$ the absolute errors (i.e. the measured minus the exact value) of these variables, respectively. If similar devices and instruments are used, it is reasonable to consider that $\varepsilon_{ah} = \varepsilon_{ac} = \varepsilon_\alpha$, that $\varepsilon_{Ph} = \varepsilon_{Pc} = \varepsilon_P$ and that $\varepsilon_{\Delta Th} = \varepsilon_{\Delta Tc} = \varepsilon_{\Delta T}$.

The probable error of $H_{QUB}$ obtained from measurements is (see Annex 2):

$$\varepsilon_{Hm} = \left[\left(\varepsilon_\alpha \frac{\partial H_{QUB}}{\partial \alpha_h}\right)^2 + \left(\varepsilon_\alpha \frac{\partial H_{QUB}}{\partial \alpha_c}\right)^2 + \left(\varepsilon_P \frac{\partial H_{QUB}}{\partial P_h}\right)^2 + \left(\varepsilon_P \frac{\partial H_{QUB}}{\partial P_c}\right)^2 \right. \tag{13}$$
$$\left. + \left(\varepsilon_{\Delta T} \frac{\partial H_{QUB}}{\partial \Delta T_h}\right)^2 + \left(\varepsilon_{\Delta T} \frac{\partial H_{QUB}}{\partial \Delta T_c}\right)^2\right]^{1/2}$$

The error coming from QUB method, equation (11), and the error coming from measurement, equation (13), are statistically independent. Then, the total error is:

$$\varepsilon_H = \sqrt{\left(H_{QUB} - H\right)^2 + \varepsilon_{Hm}^2} \tag{14}$$

and the total relative error is

$$\varepsilon_{H\%} = \frac{\sqrt{\left(H_{QUB} - H\right)^2 + \varepsilon_{Hm}^2}}{H} \tag{15}$$

A reasonable estimation of the error coming from measurement is about 0.5 – 1 °C (Strachan, Monari, Kersken, & Heusler, 2015). Therefore, modelling errors of 1 °C may be considered in the range of the measurement error, while modelling errors of 2 – 3 °C are usual (Strachan, Svehla, Heusler, & Kersken, 2016).

## 5 Global conductance of a building

The aim of QUB method is to measure the global heat conductance (or resistance) of a house. While the thermal transmittance of a building element is well defined (ISO 7345, 2018; ISO 13789, 2017), the areal heat transfer coefficient of the building, $H'$, given by equation (2), requires "a conventional definition of internal temperature, external temperature, area and the different contributions resulting in the heat flow rate. The heat flow rate may optionally include the contributions of heat transmission through building envelope, thermal bridges, ventilation, solar radiation, etc. The area may optionally be the envelope area, the floor area, etc. The temperature difference may optionally be a weighted temperature difference." (ISO 7345, 2018). Proposals and justifications for these conventional definitions are discussed hereafter.

## 5.1 One boundary temperature

The global conductance is well defined in steady-state if there is only one outdoor temperature, $T_o$, for the building (Figure 2.a):

$$H \equiv \frac{P}{\theta_i - T_o} \tag{16}$$

This is a common interpretation of the overall conductance. Equation (16) is valid for steady-state. Since steady-state can never be fully achieved for a building, methods such as degree-days or degree-hours consider that the conductance, $H \equiv UA$, is constant and estimate the global transmittance by the integral in time (Ghiaus & Allard, 2006):

$$H = \frac{\int_0^{t_{final}} P dt}{\int_0^{t_{final}} (\theta_i - T_o) dt} \tag{17}$$

In dynamic model, the conductance is the reciprocal of the static gain, which is the total resistance (Figure 2c).

A simple thermal circuit shows that the outdoor temperature, $T_o$, and the power delivered to the building, $P$, are sources (Figure 2b). In a bloc diagram, the sources are inputs (Figure 2c, d & e). For a linear model, the blocs are transfer functions which can be represented generally by $KG(s)$, where $K$ is the static gain and $G(s)$ is the dynamic gain. For strictly causal systems, the dynamic gain $G(s)$ tends to zero when time tends to infinity, $s \to 0 \Leftrightarrow t \to \infty$.

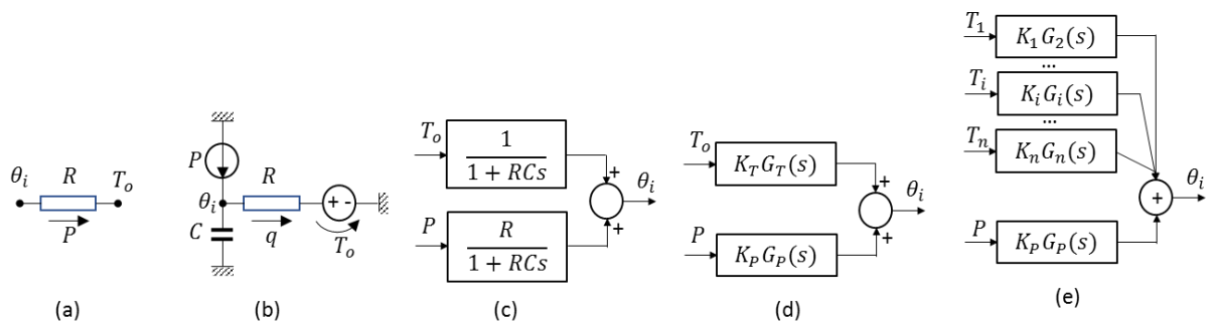

Figure 2 Definition of the total conductance: a) on a resistance; b) on a 1st order thermal circuit model; c) on a bloc diagram of a 1st order model (which corresponds to the thermal circuit from b); d) on a bloc diagram of a model with only one temperature as an input; e) on a bloc diagram of a model with more temperatures as inputs.

For a building, the static gain of the outdoor temperature is $K_T = 1$ (from physical reasons, when $P = 0$, if $t \to \infty$, then $\theta_i \to T_o$). The static gain of the power is the global resistance, $K_P = R \equiv 1/H$ (Figure 2d).

## 5.2 More boundary conditions

When there are more boundary conditions (Figure 2e), the definition is not so straightforward. The output in steady-state is:

$$\theta_i = \sum_i K_i T_i + K_p P \quad (18)$$

where $K_i$ are the static gains of the inputs that are temperatures and $K_P$ is the static gain of the power, $P$. This is true, whether all outdoor temperatures are the same, $\forall i, T_i = T_o$, or not. Note that $\sum_i K_i = 1$ (Ghiaus C. , 2013).

Therefore, by considering that the global conductance is a function of the supplied power and the temperature difference between indoor and outdoor, equation (16) will give errors in steady-state when there are multiple exterior boundary temperatures.

## 5.3 Multi-zone buildings

In the case of a multi-zone building, i.e. more indoor temperatures, the principle is to find the mean or equivalent thermal conductance. For a multi-zone building, we need the equivalent (mean) temperature, $\bar{\theta}$, and the equivalent (global) conductance, $H \equiv UA$, which produce the same heat loss over the whole surface area $A$:

$$P = UA(\bar{\theta} - T_o) \quad (19)$$

as the temperature differences $(\theta_i - T_i)$ over the surface areas $A_i$:

$$P = \sum_i U_i A_i (\theta_i - T_i) \quad (20)$$

Therefore, for determining $U$, the mean temperature $\bar{\theta}$ and the transmittances of the elements $U_i A_i$ are needed.

### 5.3.1 Mean temperature

In order to determine the total conductance, $H \equiv UA$, from equation (19), the equivalent mean temperature, $\bar{\theta}$, is needed. It is the temperature which gives the same accumulation of energy in the zone as if there were distributed temperatures:

$$\sum_i m_i c \bar{\theta} = \sum_i m_i c \theta_i \quad (21)$$

Therefore,

$$\bar{\theta} = \frac{\sum_i m_i \theta_i}{\sum_i m_i} \quad (22)$$

Simplifying by the specific mass and by the height of the zones, gives

$$\bar{\theta} = \frac{\sum_i A_i \theta_i}{\sum_i A_i} \qquad (23)$$

### 5.3.2 Total conductance obtained from zone temperatures and conductance of elements

The total conductance $H \equiv UA$ can be defined in function of the zone temperatures and the conductance of the elements. From (19) and (20) it results

$$US(\bar{\theta} - T_o) = \sum_i U_i A_i (\theta_i - T_o) \qquad (24)$$

where

$$A = \sum_i A_i \qquad (25)$$

which has two unknowns: $\bar{\theta}$ and $UA \equiv K$. Without reducing the generality, let us consider two zones for which,

$$U_1 A_1 (\theta_1 - T_o) + U_2 A_2 (\theta_2 - T_o) = U(A_1 + A_2)(\bar{\theta} - T_o) \qquad (26)$$

which yields the global conductance

$$U\bar{\theta} = \frac{U_1 A_1 \theta_1 + U_2 A_2 \theta_2}{A_1 + A_2} \qquad (27)$$

Therefore, from (16) and (27), the equivalent conductance for multi-zone is

$$H \equiv UA = \sum_i U_i A_i \theta_i \cdot \frac{\sum_i m_i}{\sum_i m_i \theta_i} \qquad (28)$$

There are problems with equation (28):
- the conductances $U_i A_i$ are difficult to obtain in the case of coupled conductive – radiative heat transfer;
- the temperatures $\theta_i$ are needed.

### 5.3.3 Total conductance obtained from state-space model

Reformulated in terms of thermal circuits, the problem of finding the equivalent total conductance, eqs. (19) and (20), becomes:

> transform the complete thermal circuit characterized in steady-state by (Ghiaus C. , 2013; Strang, 2007):
> $$0 = -\mathbf{A}^T \mathbf{G} \mathbf{A} \boldsymbol{\theta} + \mathbf{A}^T \mathbf{G} \mathbf{b} + \mathbf{f} \qquad (29)$$
> into a circuit with only one equivalent conductance (Figure 2b):
> $$0 = -G\bar{\theta} + GT_o + f \qquad (30)$$
> so that the load is the same $f = \sum_i P_i = \sum \mathbf{f}$.

If the temperature sources are zero, or if they are the same for all boundaries, i.e. $\mathbf{b} = \mathbf{0}$, then the thermal circuit (29) can be written in matrix form as

$$\mathbf{K\theta} = \mathbf{f} \tag{31}$$

where $\mathbf{K} \equiv \mathbf{A}^T \mathbf{G} \mathbf{A}$. Without reducing the generality, let us consider a two-zone building (i.e. the size of $\boldsymbol{\theta}$ is two). If there is only one outdoor temperature, then it becomes the reference temperature, $T_o = 0$. Therefore, the problem of finding the equivalent total conductance becomes:

$$\begin{bmatrix} \text{transform the complete model} \\ \mathbf{K\theta} = \mathbf{f} \; ; \begin{bmatrix} K_{11} & K_{12} \\ K_{21} & K_{22} \end{bmatrix} \begin{bmatrix} \theta_1 \\ \theta_2 \end{bmatrix} = \begin{bmatrix} f_1 \\ f_2 \end{bmatrix} \\ \text{into a model with only one conductance} \\ H\bar{\theta} = f_1 + f_2 \end{bmatrix} \tag{32}$$

$$H\bar{\theta} = f_1 + f_2 \tag{33}$$

Multiplying (32) by the line matrix

$$\begin{bmatrix} 1 & 1 \end{bmatrix} \begin{bmatrix} K_{11} & K_{12} \\ K_{21} & K_{22} \end{bmatrix} \begin{bmatrix} \theta_1 \\ \theta_2 \end{bmatrix} = \begin{bmatrix} 1 & 1 \end{bmatrix} \begin{bmatrix} f_1 \\ f_2 \end{bmatrix} \tag{34}$$

we obtain

$$(K_{11} + K_{12})\theta_1 + (K_{12} + K_{22})\theta_2 = f_1 + f_2; \tag{35}$$

or, by using equation (33)

$$(K_{11} + K_{12})\theta_1 + (K_{12} + K_{22})\theta_2 \equiv H\bar{\theta} \tag{36}$$

From the definition of $\bar{\theta}$, equation (22), and from equation (35), it results

$$H = \frac{(K_{11} + K_{12})\theta_1 + (K_{12} + K_{22})\theta_2}{\frac{m_1}{m}\theta_1 + \frac{m_2}{m}\theta_2} \tag{37}$$

The advantage of using the state-space model for calculating the global conductance is evident when the heat transfer phenomena are coupled and the coefficients $K_{ij}$ from equation (32) do not have simple expressions. For a state space model,

$$\begin{cases} \dot{\mathbf{x}} = \mathbf{Ax} + \mathbf{Bu} \\ \mathbf{y} = \mathbf{Cx} + \mathbf{Du} \end{cases} \tag{38}$$

the static gain $K_{ij}$ of the transfer function between input $j$ and output $i$ can be obtained as the steady state response to a unitary step (Rowell, 2002)

$$\mathbf{y}_{ss} = [-\mathbf{CA}^{-1}\mathbf{B} + \mathbf{D}]\mathbf{u}_{ss} \tag{39}$$

where $\mathbf{u}_{SS}$ is the vector of unitary step inputs $u_1(t), \dots, u_n(t)$:

$$\mathbf{u}_{SS} = \begin{bmatrix} u_1(t) \\ \dots \\ u_n(t) \end{bmatrix} \quad (40)$$

The static gains $K_{ij}$ between all outputs $i$ and the input $j$ is the steady-state value of the transfer function for input $j$ when $u_j = 1$ and $u_{i \neq j} = 0$. By using the solution (39) of the steady-state model and equation (22) for the mean indoor temperatures, the building heat transfer coefficient or the overall thermal conductance, $H$, is then

$$H = \frac{\sum_i P_i}{\frac{\sum_i m_i \theta_i}{\sum_i m_i} - T_o} \quad (41)$$

where
- $P_i$ is the power delivered to the air in zone $i$;
- $m_i$ – mass of air in zone $i$;
- $\theta_i$ – temperature of zone $i$ obtained from the output vector $\mathbf{y}_{ss}$ given by equation (39);
- $T_o$ – outdoor temperature.

# 6 Design of experiments for QUB method

The design of experiment is based on the calculation of the error of the QUB method, $\varepsilon_{QUB\%} = (H_{QUB} - H)/H$, as a function of the power $P_h$ and the time duration of the experiment, $t_{QUB}$ (Figure 1). Examples are given for two buildings:
1. A test building with one-thermal zone (bungalow) considering only the temperatures of the outdoor air and the ground.
2. A multi-zone building with multiple exterior conditions.

## 6.1 Obtaining the response of the model to QUB test

The values used in the definition of $H_{QUB}$ given by equations (5) or (9) are obtained from the step response of the state-space model (38) with initial conditions $\mathbf{x}(0)$ (Rowell, 2002):

$$\begin{cases} \mathbf{x}(t) = e^{\mathbf{A}t}\mathbf{x}(0) + \mathbf{A}^{-1}(e^{\mathbf{A}t} - \mathbf{I})\mathbf{B}\mathbf{u} \\ \mathbf{y}(t) = \mathbf{C}\mathbf{x} + \mathbf{D}\mathbf{u} \end{cases} \quad (42)$$

where the matrix exponential of square matrix $\mathbf{A}$ is the state transition matrix $e^{\mathbf{A}t} \equiv \mathbf{\Phi}(t) = \mathbf{V}e^{\mathbf{\Lambda}t}\mathbf{V}^{-1}$, with $\mathbf{V}$ the matrix of eigenvectors and $\mathbf{\Lambda}$ the matrix of eigenvalues of matrix $\mathbf{A}$. The initial state in equation (42) is obtained from the boundary conditions, $\mathbf{u}(0)$:

$$\mathbf{x}(0) = -\mathbf{A}^{-1}\mathbf{B}\mathbf{u}(0) \quad (43)$$

The value of the indoor temperature as a response to a step input is calculated from (42)
- at time $t_{QUB}$ by considering the initial state $\mathbf{x}(0)$ and
- at time $2t_{QUB}$ by considering the initial state as $\mathbf{x}(t_h) \equiv \mathbf{x}_{s1}$.

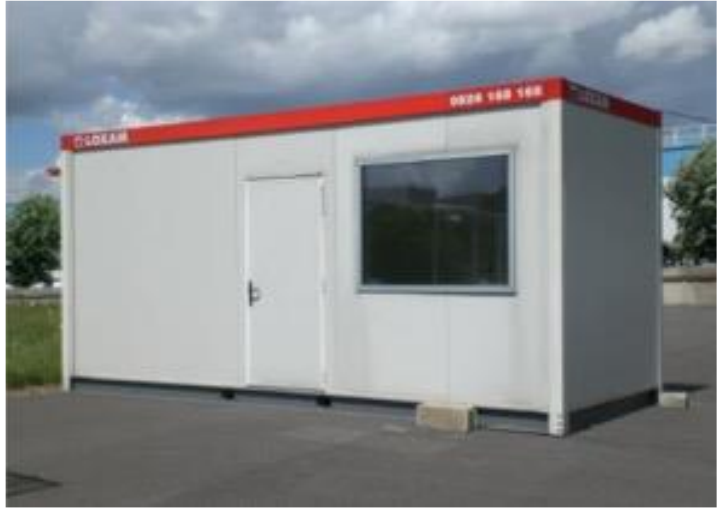
Figure 3 General view of the bungalow

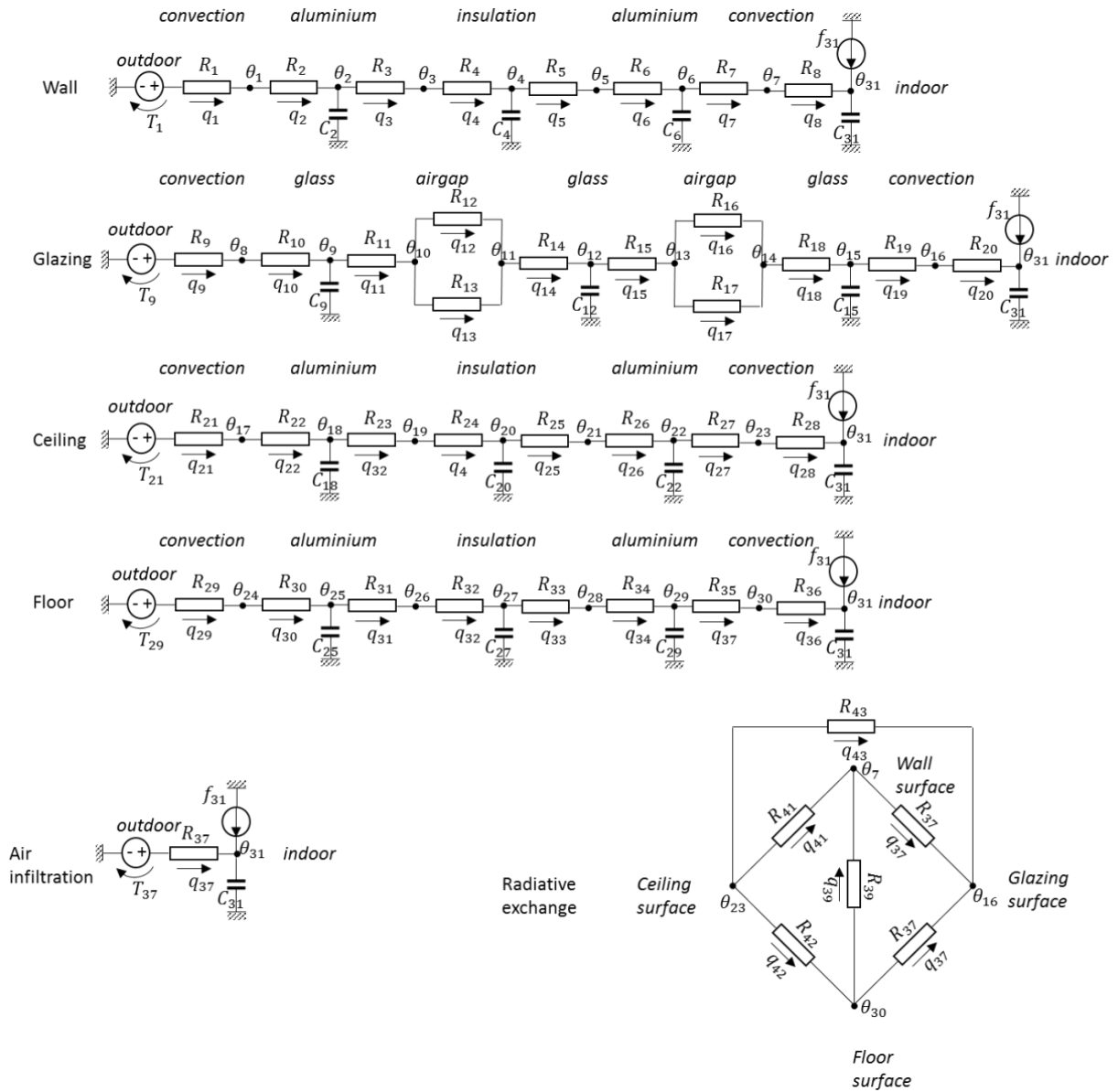
Figure 4 Thermal circuit model of the bungalow

## 6.2 Obtaining the response as a sum of exponentials

The step response given by equation (42) can be written as a function of eigenvectors $\mathbf{V}$ and eigenvalues $\mathbf{\Lambda}$ as

$$\mathbf{y}(t) = \mathbf{CM}e^{\mathbf{\Lambda} t}\mathbf{M}^{-1}\mathbf{x}(0) + (\mathbf{CA}^{-1}\mathbf{M})e^{\mathbf{\Lambda} t}(\mathbf{M}^{-1}\mathbf{Bu}) - \mathbf{CA}^{-1}\mathbf{Bu} + \mathbf{Du} \qquad (44)$$

Let us emphasize the coefficients of the exponentials by rewriting (44) as:

$$\mathbf{y}(t) = \mathbf{CM}\,\mathrm{diag}(\mathbf{M}^{-1}\mathbf{x}(0))e^{\mathbf{\Lambda} t} + \mathbf{CA}^{-1}\mathbf{M}\,\mathrm{diag}(\mathbf{M}^{-1}\mathbf{Bu})e^{\mathbf{\Lambda} t} \\ - (\mathbf{CA}^{-1}\mathbf{B} + \mathbf{D})\mathbf{u} \qquad (45)$$

The right-hand of equation (45) has three terms:
- the 1st term is the exponential response influenced by the initial conditions;
- the 2nd term is the exponential response influenced by the step input;
- the 3rd term is the steady-state value (which does not depend on the initial conditions).

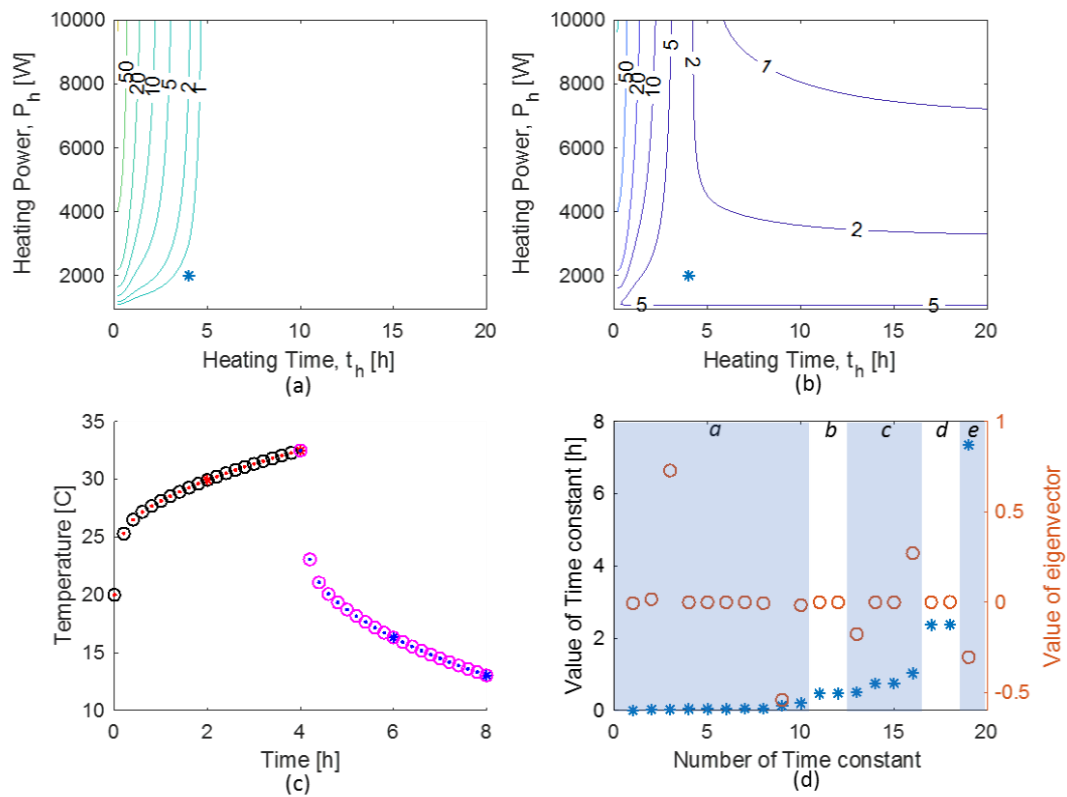

Figure 5 Design of experiment for a bungalow: a) relative intrinsic error $\varepsilon_{QUB\%}$ as a function of power and experiment duration; b) sum of intrinsic and measurement error c) time response for $t_{QUB} = 3$ h and $P_h = 2000$ W; d) Time constants (stars) and amplitudes of the corresponding exponentials (circles) of the model.

## 6.3 One-thermal zone house (bungalow)

Let us consider a single zone parallelepipedal bungalow (Figure 3). The walls are of insulated galvanized steel (surface of ceiling and ground floor 13.5 m$^2$, surface of exterior walls 37.6 m$^2$). The windows are double glazed (surface 3.88 m$^2$). The model takes into account the infiltration rate (0.5 vol/h) and the radiative exchange between the walls (Figure 4). The spatial discretization of the walls can be done arbitrarily small; the one-slice discretization shown in Figure 4 is used only for convenience of representation.

### 6.3.1 Only one boundary temperature (i.e. outdoor temperature)

The total resistance ($R \equiv 1/H$) of the model of the bungalow is determined by

$$\mathbf{y}_{ss} = [-\mathbf{C}\mathbf{A}^{-1}\mathbf{B} + \mathbf{D}]\mathbf{u} = R = 0.0191 \text{ °C/W} \tag{46}$$

for the unitary step input $\mathbf{u} = [T_O \ T_O \ T_O \ T_O \ T_O \ P]^T = [0\ 0\ 0\ 0\ 0\ 1]^T$. For our example, the largest time constant is $T = 26.4 \cdot 10^3$ s $= 7.34$ h (Figure 5d).

The results are given in Figure 5. Power larger than 943 W is needed for heating. Figure 5a shows the intrinsic error of the QUB method as a function of heating power, $P_h$, and time, $t_h$. It can be seen that for a heating time larger than 4 h, the intrinsic error of the method is zero for any power (Figure 5a). However, measurements are affected by errors. Figure 5b shows the estimated error obtained by adding the measurement error given by equation (74). It can be noticed that important errors result by diminishing the heating time and the heating power. In order to obtain robust estimations, the heating time needs to be larger than 3 hours and the power larger than about 2000 W. The value of the heat transfer coefficient found by the QUB experiment is $H_{QUB} = 61.2$ W/K which can be compared with the initial supposed value $H_{QUB} = 52.4$ W/K obtained from the model used for the design of the experiment (equation (46)). The values of the heat transfer coefficient found by QUB experiment was compared with the value found by co-heating; as in other experiments, the differences between QUB and co-heating were about 5 – 10 % (Mangematin, Pandraud, & Roux, 2012; Meulemans J. , 2018; Alzetto, Farmer, Fitton, Huges, & Swan, 2018; Alzetto, Pandraud, Fitton, & Heusler, 2018; Alzetto, Gossard, & Pandraud, 2014)

The fact that small measurement errors can be obtained for relatively short heating duration can be explained by the time constants and the coefficients of the exponential response of the model given by equation (45) and shown in Figure 5d. For the model of the bungalow, the time constants can be grouped in five classes:
  a) Very short time constants ($< 0.1$ h) with significant coefficients in equation (45). Their response is so quick that they act as a step response at the time scale of the heating and cooling.
  b) Small time constants (0.5 h ... 1 h) but with very small coefficients in equation (45). Due to the value of their coefficients, these exponentials do not have a significant contribution to the response.
  c) Medium time constants (1 ... 2 h) with significant coefficients in equation (45). These are the time constants that have a significant contribution to the exponential response in the QUB method.

d) Medium time constants (2 … 3 h) but small coefficients in equation (45). Due to the values of their coefficients, these time constants have small influence on the response.
e) Large time constants ($> 6$ h) with large coefficients in equation (45). These time constants have little influence on the response due to the relatively short heating and cooling time used in the experiment.

Therefore, the response of the system may be considered a pure exponential plus a constant when the durations of heating and cooling are larger than 3 h (4 times the time constant of 0.8 h). This observation shows an important feature of the QUB method: it allows measuring the heat transfer coefficient in a short time but this time needs to be long enough in order to obtain the steady state for the small time-constants which have significant coefficients and short enough to have small values of the exponential corresponding to large time constants. Figure 5d shows another interesting feature of the thermal models of buildings. If the model is obtained by a fine discretisation, the time constants are grouped: many small-time constants and a few, rather clearly separated, larger time constants. Figure 5c shows the time response for the conditions of the experiment. Two features can be noticed: the fast response due to the short time constants and the exponential response due to the medium time constants associated with significant coefficients.

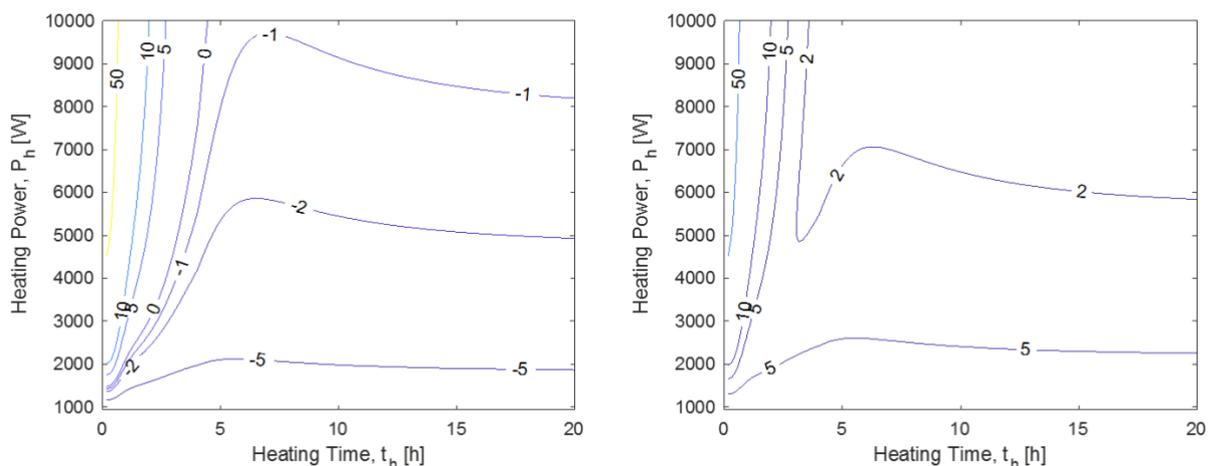

Figure 6 Error evaluation of heat transfer coefficient for the bungalow when: a) the global conductance, $K_P$, is considered; b) the experimentally measurable global conductance, $H$, is considered.

### 6.3.2 More boundary temperatures for the bungalow

Usually, the residential buildings have a non-adiabatic ground floor. Inherently, the QUB method, as any other measurement method, will do an error if the ground temperature is not taken into account. Figure 6a shows the intrinsic error of the QUB method when the value of the conductance is the correct one, i.e. representing the static gain of power calculated from the physical values; in this case, the value of $H$ does not depend on the boundary temperatures. Figure 6b shows the errors when the total conductance is obtained with equations (16) and/or (17) (i.e. by considering only the outdoor air temperature), which is a common interpretation of the total conductance; this time, the value of $H$ depends of the boundary temperatures. Figure 6 shows that, by increasing the power, the error is reduced, i.e. the influence of the ground floor in the response is reduced.

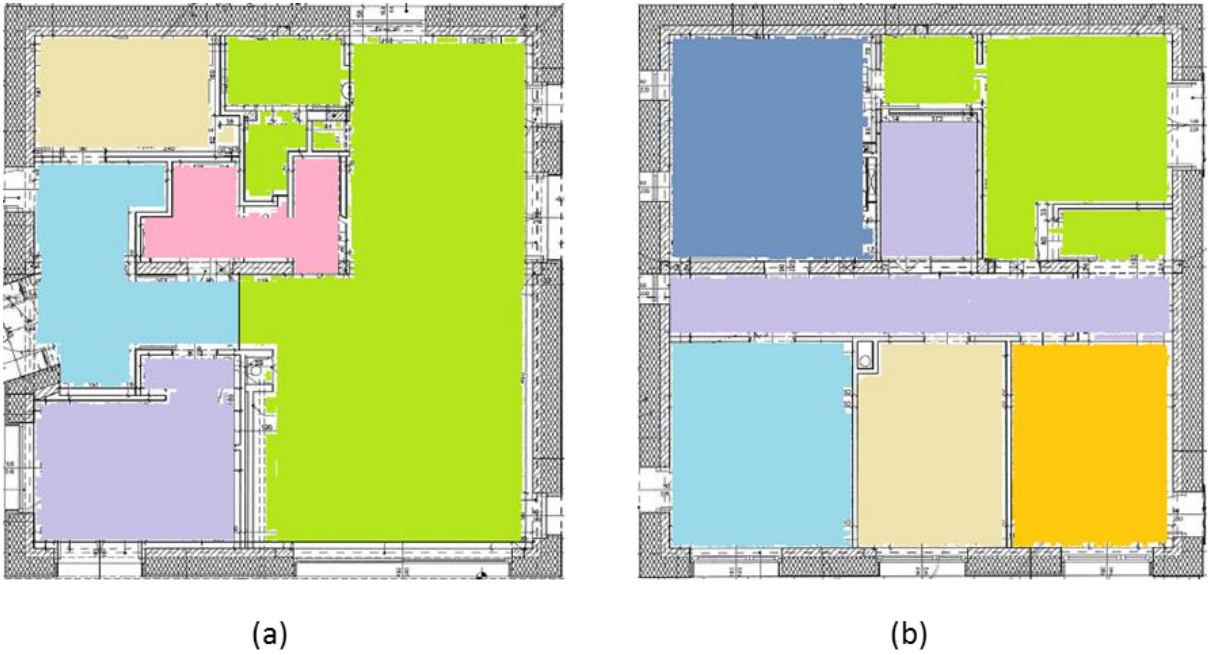

(a)                (b)

Figure 7: Thermal zones of the house considered in the experiments: (a) ground floor (b) first floor.

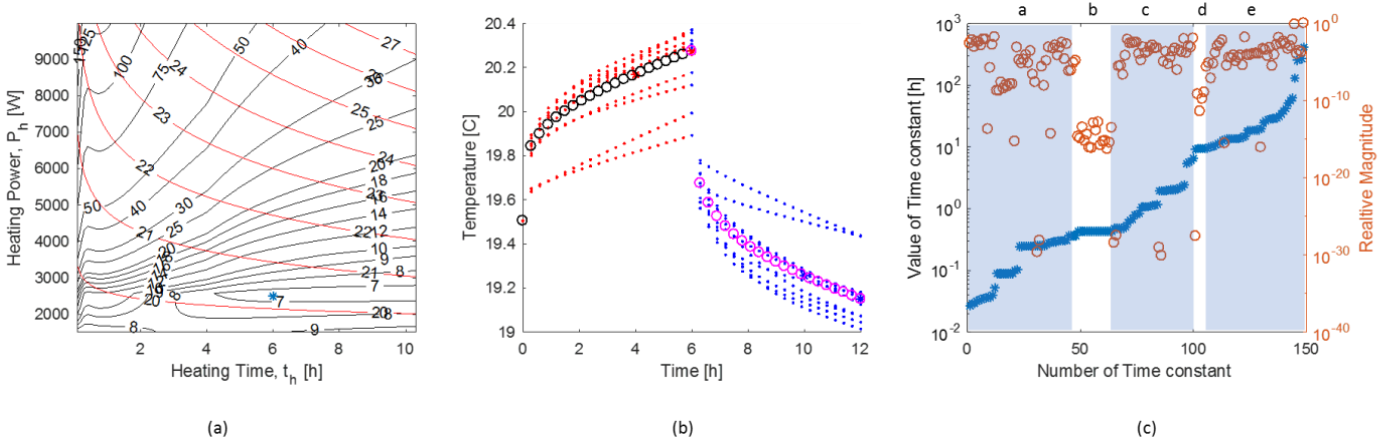

(a)           (b)           (c)

Figure 8 Design of experiment for a house: a) relative error $\varepsilon_{QUB\%}$ as a function of power and duration of experiment; b) time response for $P_h = 2500$ W and $t_{QUB} = 6$ h; c) Time constants (stars) and amplitudes of the corresponding exponentials (circles) of the model.

## 6.4 Case of a typical house

Finally, let us consider a typical two-floor house with the ground floor area of 93.3 m², the total floor area 186.6 m² and total volume of 502 m³ (Figure 7). There are triple glazed windows with aluminum frames; external walls are made of silicate bricks of 18 cm with external mineral wool insulation. The ceiling has an external insulation made of 54 cm of mineral wool. There is no crawl space below the ground floor; there are 31 cm of expanded polystyrene and 25 cm of reinforced concrete. The value of the air leakage rate measured by blower door is $n_{50} = 0.34$ vol/h at 50 Pa.

It can be seen from Figure 8a that, if the outdoor temperature is 19.5°C, the optimal design of the experiment is for a power of about 2500 W and for a heating time of 6 h. The error obtained in this case is about 7 % (coming from the ground floor) and the equivalent indoor temperature is 20.3°C. Figure 8a shows the valley in which the optimum is located. The experiment is very sensitive to power variation and less sensitive to time duration. Figure 8b shows that the response is quasi-instantaneous at the beginning (corresponding to the small time-constants), followed by an exponential form (corresponding to the medium time constants). This is explained by the distribution of the 149 time-constants and the amplitude of their corresponding coefficients obtained by using equation (45) (Figure 8c). Similarly to the case of the bungalow, the time constants can be grouped in five classes:

a) Very short time constants ($0.5 \cdot 10^{-1}$ h … $2 \cdot 10^{-1}$ h) with significant coefficients in equation (45). Their response is so quick that they act almost as a step response at the time scale of the heating and cooling.
b) Small time constants ($2 \cdot 10^{-1}$ h … $10^0$ h) but with very small coefficients in equation (45). Due to the value of the eigenvectors, they do not have a significant contribution to the response.
c) Medium time constants (around $10^0$ h) with significant coefficients in equation (45). These are the time constants that have a significant contribution to the exponential response in the QUB method.
d) Medium time constants (around $10^1$ h) but small coefficients in equation (45). Due to the values of their coefficients, these time constants have small influence on the response.
e) Large time constants ($> 10^1$ h) and large coefficients in equation (45). These time constants have little influence on the response due to the relatively short heating and cooling time.

Therefore, at the time scale of the QUB experiment, the system with these time constants gives an exponential response (Figure 8b). The distribution of time constants and of their corresponding coefficients justifies the use in QUB experiment of a short time as compared with the largest time constants of the building.

# 7  Conclusions

A method to design QUB experiments was developed and experimentally tested on two real buildings: a bungalow and a typical house. The method gives the error of the overall thermal transmittance, $H$, as a function of the heating power and the heating duration. Initial conditions, outdoor temperature and power during cooling may be introduced in the design of the experiment. The method is valid if the outdoor conditions change insignificantly during the experiment. Therefore, the experiment is conducted during the night time.

The relatively small-time duration of QUB experiment is explained by the distribution of the time constants. There are a lot of time constants that have small values, a few that have medium values and significant coefficients, and a few with large values and significant coefficients. The QUB method takes advantage of this distribution of the time constants by designing an experiment at the time scale of the medium time constants (about a few hours). Therefore, a good design of the experiment of QUB method allows to obtain measurement errors less than 10 % for real in-situ houses with experimental durations shorter than 12 h.

This method allows the experimenter to select the best input parameters (heating power and duration) based on a priori knowledge of the building envelope to be measured and on the weather forecast. It could also be used as a decision tool to make or not a measurement depending on the predicted error that would be committed as a function of building constructive characteristics and weather. This method could help improving the data analysis by selecting best period for fitting the thermal model on the measured data.

# Annexes
## Annex 1: Slope of the response curve of QUB test in heating and cooling

In heating, it is supposed that the space is submitted to two inputs: the outdoor temperature, $T_o$, and the power, $P_h$. In order to emphasize this, by considering $\Delta T \equiv \theta_i - T_o$, let's write equation (4) as

$$C\dot{\theta}_i = -H_{QUB}\theta_i + P_h + H_{QUB}T_o . \tag{47}$$

Equation (47) is linear; it can be solved by applying the superposition of two equations:

$$C\dot{\theta}_i = -H_{QUB}\,\theta_i + P_h \tag{48}$$

and

$$C\dot{\theta}_i = -H_{QUB}\,\theta_i + H_{QUB}\,T_0^h . \tag{49}$$

The general solution of equation (48) is:
$$\theta_i = c_1 \exp\left(-\frac{t}{\tau}\right) + \frac{P_h}{H_{QUB}}\,; \tau \equiv C/H_{QUB} \tag{50}$$

From the initial zero condition: $\Delta T_0^h \equiv \theta_i(0) - T_0^h$, i.e. $c_1 + \frac{P}{H'_{QUB}} = 0$, the constant $c_1 = -\frac{P}{K_0}$ is obtained. The solution of equation (48) is:

$$\theta_i = \frac{P_h}{H_{QUB}}\left(1 - \exp\left(-\frac{t}{\tau}\right)\right) \tag{51}$$

The solution of equation (49) is:

$$\begin{aligned}&C\dot{\theta}_i = -H'_{QUB}(\theta_i - T_0^h)\,; \Delta T \equiv \theta_i - T_0^h ,\\ &\Delta T = c_1 \exp(-t/\tau)\,; \Delta T(0) = \Delta T_0 ,\\ &\Delta T = \Delta T_0 \exp(-t/\tau) .\end{aligned} \tag{52}$$

The superposition gives the solution:

$$\Delta T^h = \Delta T_0^h \exp\left(-\frac{t}{\tau}\right) + \frac{P_h}{H_{QUB}}\left(1 - \exp\left(-\frac{t}{\tau}\right)\right) \tag{53}$$

The equation of the tangent in origin of the function (53) is

$$\Delta T^h = \Delta T_0^h + (P_h/C - \Delta T_0^h/\tau)t ,\tag{54}$$

with the slope

$$\alpha_h = \frac{P_h}{C} - \frac{\Delta T_0^h}{\tau} = \frac{P_h}{C} - \frac{\Delta T_0^h H_{QUB}}{C} \tag{55}$$

which is the slope in origin of the sum of two exponential responses given by equation (53).

In QUB method, the tangent to the exponential response at a time $t = t_0 > 0$ is used. The equation of the tangent at $t = t_0$ is:

$$\Delta T^h = \left(1 - e^{-\frac{t_0}{\tau}}\right)\frac{P_h}{H_{QUB}} + e^{-\frac{t_0}{\tau}}[\Delta T_0^h + (P_h/C - \Delta T_0^h/\tau)(t - t_0)] \tag{56}$$

with the slope

$$\alpha_h = e^{-\frac{t_0}{\tau}}(P/C - \Delta T_0^h H_{QUB}/C) \tag{57}$$

Assuming that the outdoor temperature, $T_o$, is constant and defining:
$\Delta T^c = \theta_i - T_o$ the difference between the indoor and outdoor temperatures;
$\Delta T_0^c = \theta_i|_{t=0} - T_o$ the difference at initial time, $t = 0$,
equation (4) becomes

$$C\frac{d}{dt}\Delta T^c = -H_{QUB}\Delta T^c + P_c \tag{58}$$

with initial condition $\Delta T^c = \Delta T_0^c$. The general solution is

$$\Delta T^c = c_1 \exp\left(-\frac{t}{\tau}\right) + \frac{P_c}{H_{QUB}} \tag{59}$$

with the constants resulting from the initial condition:

$$\Delta T^c = \Delta T_0^c e^{-\frac{t}{\tau}} + \frac{P_c}{H_{QUB}}\left(1 - e^{-t/\tau}\right); \tau \equiv C/H_{QUB} \tag{60}$$

The tangent in the origin of the exponential given by eq. (60) has the equation

$$\Delta T^c = \Delta T_0^c + \left(\frac{P_c}{H_{QUB}} - \frac{\Delta T_0^c}{\tau}\right)t \tag{61}$$

with the slope

$$\alpha_c = \frac{P_c}{H_{QUB}} - \Delta T_0^c/\tau = \frac{P_c}{H_{QUB}} - \frac{\Delta T_0^c H_{QUB}}{C} \tag{62}$$

The equation of the tangent at $t = t_0$ is

$$\Delta T^c = \left(1 - e^{-\frac{t_0}{\tau}}\right)\frac{P_c}{H_{QUB}} + e^{-\frac{t_0}{\tau}}\left[\Delta T_0^c + \left(\frac{P_c}{C} - \frac{\Delta T_0^c}{\tau}\right)(t - t_0)\right] \tag{63}$$

with the slope

$$\alpha_c = e^{-t_0 \frac{H_{QUB}}{C}} \left(\frac{P_c}{C} - \frac{\Delta T_0^c H_{QUB}}{C}\right) \tag{64}$$

## Annex 2: Probable error from measurements

The true value of $H_{QUB}$ is

$$\begin{aligned}H_{QUB}\left(\bar{\alpha}_h, \bar{\alpha}_c, \bar{P}_h, \bar{P}_c, \Delta\bar{T}_h, \Delta\bar{T}_c\right) = \\ H_{QUB}(\alpha_h - \varepsilon_\alpha, \alpha_c - \varepsilon_\alpha, P_h - \varepsilon_P, P_c - \varepsilon_P, \Delta T_h - \varepsilon_{\Delta T}, \Delta T_c - \varepsilon_{\Delta T})\end{aligned} \tag{65}$$

By developing the right-hand side in a series expression, we obtain

$$\begin{aligned}\bar{H}_{QUB} = \\ H_{QUB} - \varepsilon_\alpha \frac{\partial H_{QUB}}{\partial \alpha_h} - \varepsilon_\alpha \frac{\partial H_{QUB}}{\partial \alpha_c} - \varepsilon_P \frac{\partial H_{QUB}}{\partial P_h} - \varepsilon_P \frac{\partial H_{QUB}}{\partial P_c} - \varepsilon_{\Delta T} \frac{\partial H_{QUB}}{\partial \Delta T_h} - \varepsilon_{\Delta T} \frac{\partial H_{QUB}}{\partial \Delta T_c}\end{aligned} \tag{66}$$

where $\bar{H}_{QUB} \equiv H_{QUB}(\bar{\alpha}_h, \bar{\alpha}_c, \bar{P}_h, \bar{P}_c, \Delta\bar{T}_h, \Delta\bar{T}_c)$ and $H_{QUB} \equiv H_{QUB}(\alpha_h, \alpha_c, P_h, P_c, \Delta T_h, \Delta T_c)$

and the partial derivatives are

$$\frac{\partial H_{QUB}}{\partial \alpha_h} = \frac{\Delta T_c(P_h \alpha_c - P_c \alpha_h)}{\sigma^2} - \frac{P_c}{\sigma} \tag{67}$$

$$\frac{\partial H_{QUB}}{\partial \alpha_c} = -\frac{\Delta T_h(P_h \alpha_c - P_c \alpha_h)}{\sigma^2} + \frac{P_h}{\sigma} \tag{68}$$

$$\frac{\partial H_{QUB}}{\partial P_h} = \frac{\alpha_c}{\sigma} \tag{69}$$

$$\frac{\partial H_{QUB}}{\partial P_c} = -\frac{\alpha_h}{\sigma} \tag{70}$$

$$\frac{\partial H_{QUB}}{\partial T_h} = -\frac{\alpha_c(P_h\alpha_c - P_c\alpha_h)}{\sigma^2} \qquad (71)$$

$$\frac{\partial H_{QUB}}{\partial T_c} = \frac{\alpha_h(P_h\alpha_c - P_c\alpha_h)}{\sigma^2} \qquad (72)$$

where

$$\sigma = \Delta T_h \alpha_c - \Delta T_h \alpha_c \qquad (73)$$

Then, the probable error of the $H_{QUB}$ obtained from measurements is

$$\varepsilon_{Hm} = \left[\left(\varepsilon_\alpha \frac{\partial H_{QUB}}{\partial \alpha_h}\right)^2 + \left(\varepsilon_\alpha \frac{\partial H_{QUB}}{\partial \alpha_c}\right)^2 + \left(\varepsilon_P \frac{\partial H_{QUB}}{\partial P_h}\right)^2 + \left(\varepsilon_P \frac{\partial H_{QUB}}{\partial P_c}\right)^2 \right.$$
$$\left. + \left(\varepsilon_{\Delta T} \frac{\partial H_{QUB}}{\partial \Delta T_h}\right)^2 + \left(\varepsilon_{\Delta T} \frac{\partial H_{QUB}}{\partial \Delta T_c}\right)^2 \right]^{1/2} \qquad (74)$$